\documentclass{article}
\usepackage{frascatiphys,here,graphicx,subfigure}
\begin{document}
\title{ The WA3 data and the two $K_1(1270)$ resonances}
\author{
L. S. Geng and E. Oset\\
{\em Departamento de Fisica Teorica e IFIC,}\\
{\em Centro Mixto
Universidad de Valencia, E-46071 Valencia, Spain} \\
L. Roca and J. A. Oller\\
{\em  Departamento de Fisica. Universidad de
Murcia. E-30071 Murcia.  Spain.}
}

\maketitle

\baselineskip=11.6pt
\begin{abstract}
Recent studies based on unitary chiral perturbation theory (U$\chi$PT) found that the
low-lying axial vector mesons can be dynamically generated
due to the interaction of the pseudoscalar octet of the pion and the vector nonet of the rho.
In particular, two poles in the second Riemann sheet have been
associated to the nominal $K_1(1270)$ resonance. In this talk, we present a
recent analysis of the  WA3 data on $K^-
p\rightarrow K^-\pi^+\pi^- p$ at 63 GeV using the U$\chi$PT amplitudes, and show that it is in favor of
the existence of two $K_1(1270)$'s [Phys. Rev. D 75, 014017 (2007)]. 
\end{abstract}
\baselineskip=14pt

\section{Introduction}

The unitary extension of chiral perturbation theory, U$\chi$PT, has been
successfully applied to study many meson-baryon and meson-meson
interactions. More recently, it has been used to study the lowest
axial vector mesons $b_1(1235)$, $h_1(1170)$, $h_1(1380)$,
$a_1(1260)$, $f_1(1285)$, $K_1(1270)$ and
$K_1(1400)$~\cite{Lutz:2003fm,Roca:2005nm}. Both works generate most
of the low-lying axial vector mesons dynamically. However, there is
a surprising discovery in Ref.~\cite{Roca:2005nm}, i.e., two poles are found
in the second Riemann sheet in the $S=1$ and $I=1/2$ channel and both are attributed to the $K_1(1270)$.

 Although the $K_1(1270)$ has been observed in various reactions,  the most conclusive
 and high-statistics data of the $K_1(1270)$ come from the WA3 experiment
 at CERN that accumulated data on the reaction $K^-p\to K^- \pi^+ \pi^- p$
 at 63 GeV. These data were analyzed by the ACCMOR
 Collaboration~\cite{Daum:1981hb}.
 As will be shown in this paper, the two-peak structure, with a peak
 at lower energy depending drastically on the reaction channel
 investigated,
 can be easily explained in our model with two poles for the
 $K_1(1270)$ plus the $K_1(1400)$. With only one pole, as has been noted long time
 ago~\cite{Daum:1981hb,Bowler:1976qe}, there is always a discrepancy for the
 peak positions observed in the $K^*\pi$ and $\rho K$ invariant mass
 distributions.

\section{Chiral unitary model and the two $K_1(1270)$'s}
In the following, we briefly describe the chiral unitary approach, while detailed formalism can be found in
 Refs.~\cite{Geng:2006yb,Roca:2005nm}.  In the Bethe-Salpeter formulation of the unitary chiral perturbation theory~\cite{Oller:1997ti},
 one has
the following unitarized amplitude:
\begin{equation}
T=[1+V\hat{G}]^{-1}(-V) \,\vec{\epsilon}\cdot\vec{\epsilon}\,',
\label{bethes}
\end{equation}

\noindent where $\hat{G}=(1+\frac{1}{3}\frac{q^2_l}{M_l^2})G$ is a
diagonal matrix with the $l-$th element, $G_l$, being the two meson
loop function containing a vector and a pseudoscalar meson:
\begin{equation}
G_{l}(\sqrt{s})= i \, \int \frac{d^4 q}{(2 \pi)^4} \,
\frac{1}{(P-q)^2 - M_l^2 + i \epsilon} \,
 \frac{1}{q^2 - m^2_l + i
\epsilon}, \label{loop}
\end{equation}
\noindent with $P$ the total incident momentum, which in the center
of mass frame is $(\sqrt{s},0,0,0)$. The loop function $G_l$ can be regularized either
by a cutoff or by dimensional regularization. In the former case, one has cutoff values, whereas in the latter, one
has subtraction constants as free parameters, which have to be fitted to the data.
\begin{figure}[H]
\begin{center}
\includegraphics[scale=0.27]{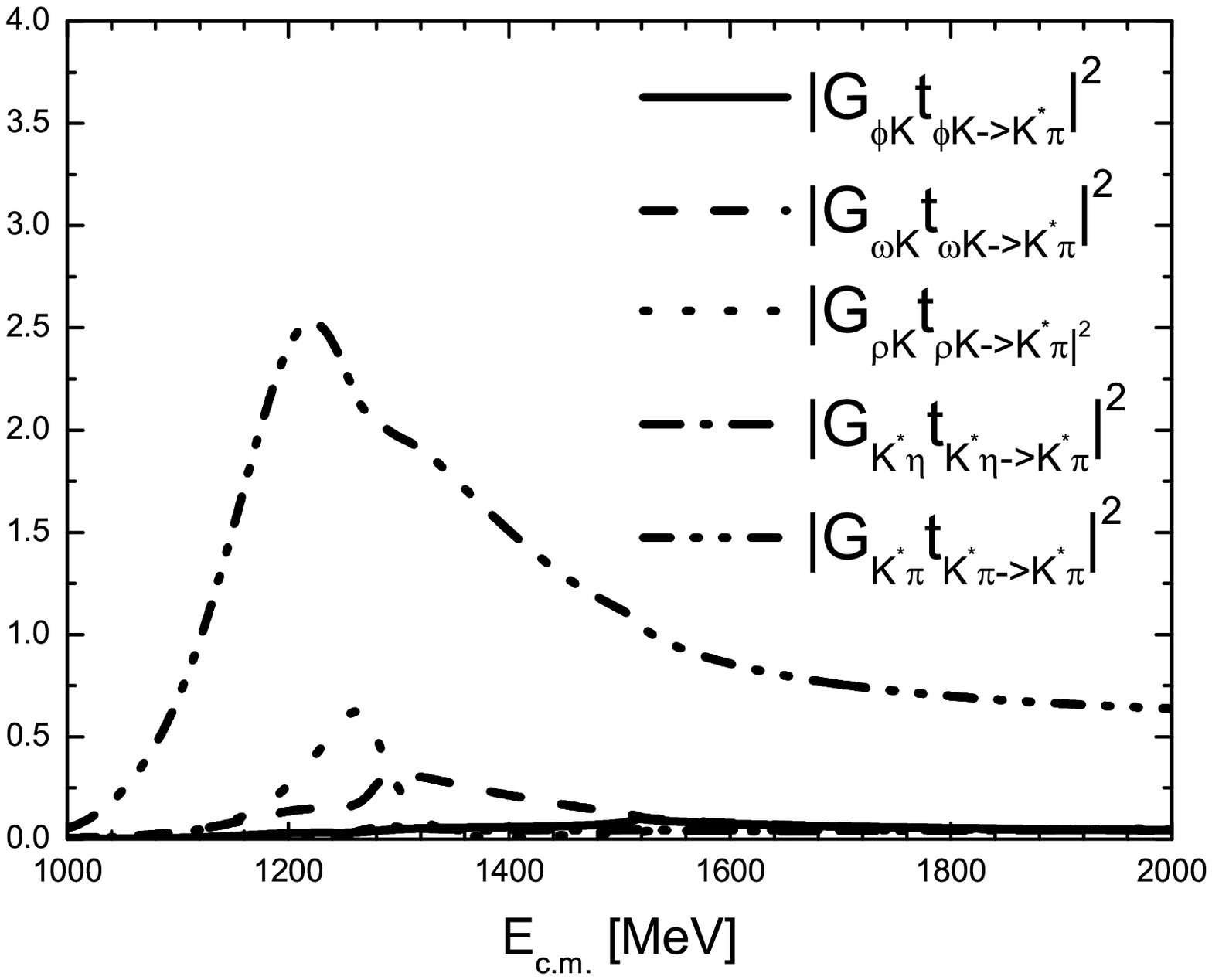}%
\includegraphics[scale=0.27]{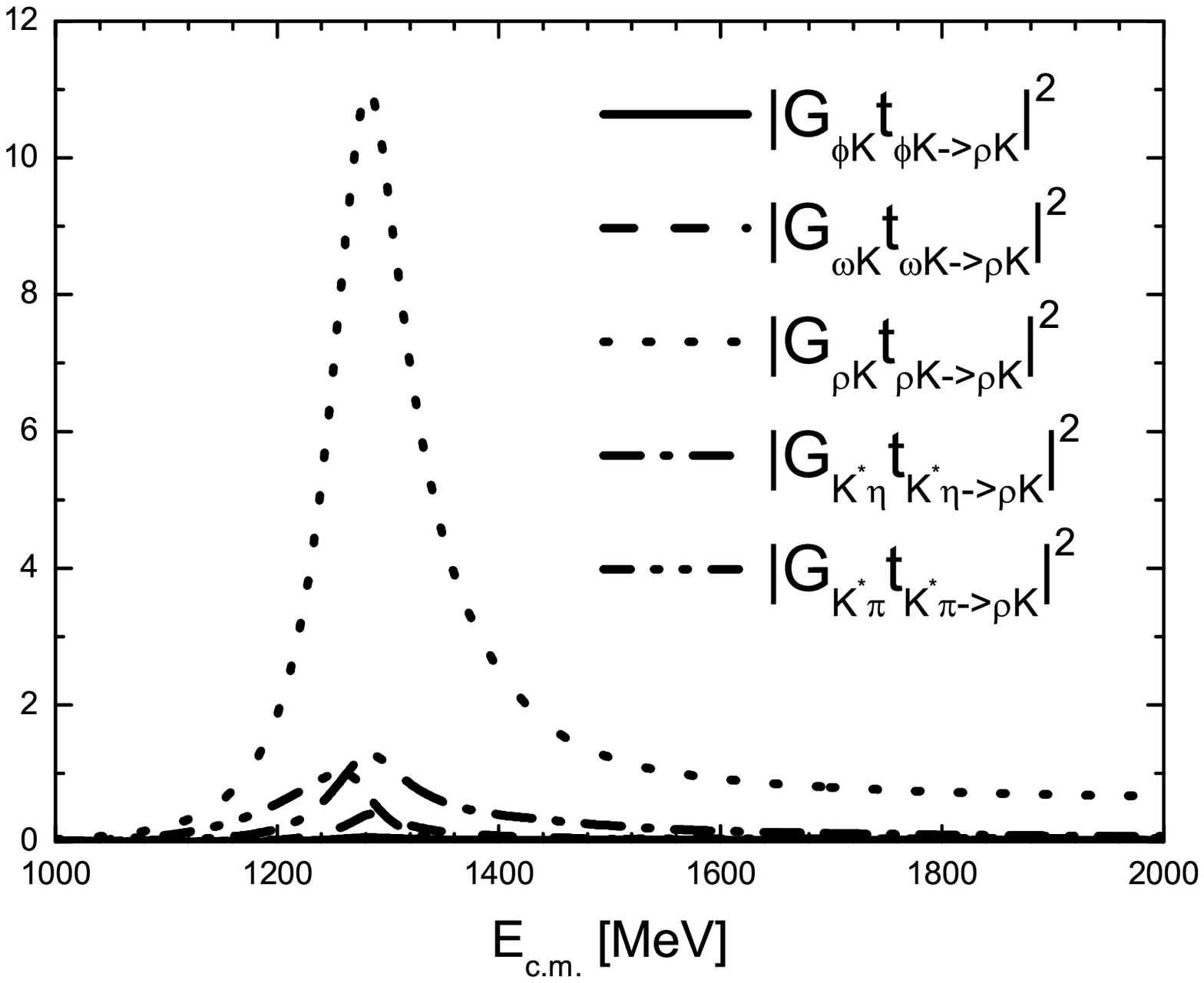}
\caption{The modulus squared of the coupled channel amplitudes
multiplied by the corresponding loop functions in the $S=1$ and
$I=\frac{1}{2}$ channel.}
\label{fig:GtVPVP}%fig2
\end{center}
\end{figure}

The tree level amplitudes are calculated using the following
interaction Lagrangian~\cite{Birse:1996hd}:
\begin{equation}
{\cal L}_{I}=-\frac{1}{4}\mathrm{Tr}\left\{\left(\nabla_\mu V_\nu-
\nabla_\nu V_\mu\right) \left(\nabla^\mu V^\nu-\nabla^\nu
V^\mu\right)\right\}, \label{eq:LBirse}
\end{equation}
\noindent where $\mathrm{Tr}$ means SU(3) trace and $\nabla_\mu$ is
the covariant derivative defined as
\begin{equation}
\nabla_\mu V_\nu=\partial_\mu V_\nu+[\Gamma_\mu , V_\nu],
\end{equation}
\noindent where $[,]$ stands for commutator and $\Gamma_\mu$ is the
vector current $
\Gamma_\mu=\frac{1}{2} (u^\dagger\partial_\mu u +u\partial_\mu
u^\dagger) $ with $
u^2=U=e^{i\frac{\sqrt{2}}{f}P}. \label{eq:ufields} $
In the above equations $f$ is the pion decay constant in the chiral
limit and $P$ and $V$ are the SU(3) matrices containing the pseudoscalar octet of the pion
 and the vector nonet of the rho.

 Fig. \ref{fig:GtVPVP} shows the modulus squared of the $S=1$, $I=\frac{1}{2}$ amplitudes multiplied
by the corresponding loop functions obtained with $f=115$ MeV,
$a(\mu)=-1.85$ and $\mu=900$ MeV. The pole positions and
corresponding widths obtained with this set of parameters are shown in Table \ref{tab:poles2}.
From Fig.~\ref{fig:GtVPVP},
 the two poles are clearly seen:  the higher pole
manifests itself as one relatively narrower resonance around
1.28 GeV and the lower pole as a broader resonance at
$\sim1.20$ GeV.

 The effective couplings
for the coupled channels $\phi K$, $\omega K$, $\rho K$, $K^*\eta$ and $K^*\pi$, calculated from the residues of the
amplitudes at the complex pole positions, are
tabulated in Table~\ref{tab:coup}
 for both the lower pole and the higher pole,
respectively.
\begin{table}[t]
    \setlength{\tabcolsep}{0.3cm}
    \renewcommand{\arraystretch}{1.2}
\caption{Effective couplings of the two poles of the $K_1(1270)$ to the
five channels: $\phi K$, $\omega K$, $\rho K$, $K^*\eta$ and
$K^*\pi$. All the units are in MeV.\label{tab:poles2}}
\begin{center}
\begin{tabular}{c|cc|cc}
\hline\hline &&&& \\[-4.5mm]
 $\sqrt{s_p}$ & \multicolumn{2}{c|}{$1195-i123$} &\multicolumn{2}{c}{ $1284-i73$ } \\
\cline{2-5}
& $g_i$ & $|g_i|$ & $g_i$ & $|g_i|$ \\
\hline  &&&& \\[-4.5mm]
$\phi K$   &  $2096-i1208$  & $2420$ & $1166-i774$  & $1399$  \\
$\omega K$ &  $-2046+i821$ & $2205$ & $-1051+i620$ & $1220$  \\
$\rho K$   &  $-1671+i1599$& $2313$ & $4804+i395$  & $4821$  \\
$K^* \eta$ &  $72+i197$    & $210$  & $3486-i536$   & $3526$  \\
$K^* \pi$  &  $4747-i2874$ & $5550$ & $769-i1171$   & $1401$  \\
\hline\hline
\end{tabular}
\label{tab:coup}
\end{center}
\end{table}
It is clearly seen that the lower pole couples
dominantly to the $K^*\pi$ channel while the higher pole couples
more strongly to the $\rho K$ channel. If different
reaction mechanisms favor one or the other channel, they will see
different shapes for the resonance. More importantly, it is to be
noted that not only the two poles couple to different channels with
different strengths, but also they manifest themselves in different
final states. In other words, in the $\rho K$ final states, one
favors a narrower resonance around 1.28 GeV, while in the $K^*\pi$
final states, one would favor a broader resonance at a smaller
invariant mass.
\section{Studying the WA3 data with the U$\chi$PT amplitudes}
 The reaction $K^-p\rightarrow K^-\pi^+\pi^- p$ can be
analyzed by the isobar model as $K^-p\rightarrow
(\bar{K}^{*0}\pi^-\,\mbox{or}\,\rho^0 K^-) p\rightarrow
K^-\pi^+\pi^- p$. Therefore, one can construct the following
amplitudes to simulate this process. Assuming $I=\frac{1}{2}$
dominance for $\bar{K}^{*0}\pi^-$ and $\rho^0 K^-$ as suggested by
the experiment one has
%\begin{widetext}
\begin{eqnarray}
T_{K^*\pi}\equiv
T_{\bar{K}^{*0}\pi^-}&=&\sqrt{\frac{2}{3}}a+\sqrt{\frac{2}{3}}aG_{K^*\pi}t_{K^*\pi\rightarrow
K^*\pi}+\sqrt{\frac{2}{3}}bG_{\rho
K}t_{\rho K\rightarrow K^*\pi},\nonumber \\
T_{\rho K}\equiv T_{\rho^0
K^-}&=&-\sqrt{\frac{1}{3}}b-\sqrt{\frac{1}{3}}aG_{K^*\pi}t_{K^*\pi\rightarrow
\rho K}-\sqrt{\frac{1}{3}}bG_{\rho K}t_{\rho K\rightarrow \rho K},
\label{eq:Ts}
\end{eqnarray}
%\end{widetext} 
where $t_{ij}$ are the coupled channel amplitudes
obtained in Section 2 and the Clebsch-Gordan coefficient
$\sqrt{\frac{2}{3}}$($-\sqrt{\frac{1}{3}}$) accounts for projecting
the $I=\frac{1}{2}$ $K^*\pi$ ($\rho K$) state into
$\bar{K}^{*0}\pi^-$($\rho^0 K^-$). The coefficients $a$ and $b$ are
complex couplings. 

To contrast our model with data, it is necessary  to take into
account the existence of the $K_1(1400)$, which is not dynamically
generated in our approach. Therefore, we add to the amplitudes in
Eq.~(\ref{eq:Ts}) an explicit contribution of the $K_1(1400)$
 \begin{eqnarray}
 T_{K^*\pi}&\rightarrow&
 T_{K^*\pi}+\frac{g_{K^*\pi}}{s-M^2+iM\Gamma(s)},\nonumber\\
  T_{\rho K}&\rightarrow&
 T_{\rho K}+\frac{g_{\rho K}}{s-M^2+iM\Gamma(s)},
 \label{bwadd}
 \end{eqnarray}
  where $g_{K*\pi}$ and $g_{\rho K}$ are complex couplings, and $M$ and $\Gamma(s)$ are the mass and width of the $K_1(1400)$ with
the $s$-wave width given by
\begin{equation}
\Gamma(s)=\Gamma_0\frac{q(s)}{q_\mathrm{on}}\Theta(\sqrt{s}-M_{K^*}-M_{\pi}).
\end{equation}
$q(s)$ and $q_\mathrm{on}$ are calculated by
 \begin{equation}
 q(s)=\frac{\lambda^{1/2}(s,M^2_\pi,M^2_{K^*})}{2\sqrt{s}}\quad\mbox{and}\quad
 q_\mathrm{on}=\frac{\lambda^{1/2}(M^2,M^2_\pi,M^2_{K^*})}{2M}.
 \end{equation}

In our model, Eq.~(\ref{bwadd}), we have the following adjustable
parameters: $a$, $b$, $g_{K^*\pi}$, $g_{\rho K}$, $M$ and
$\Gamma_0$. In principle, $f$ and $a(\mu)$ can also be taken as free
parameters. One can then fix these parameters by fitting the WA3 data (see Ref.~\cite{Geng:2006yb} for details).
\begin{figure}[htpb]
\begin{center}
\includegraphics[scale=0.27]{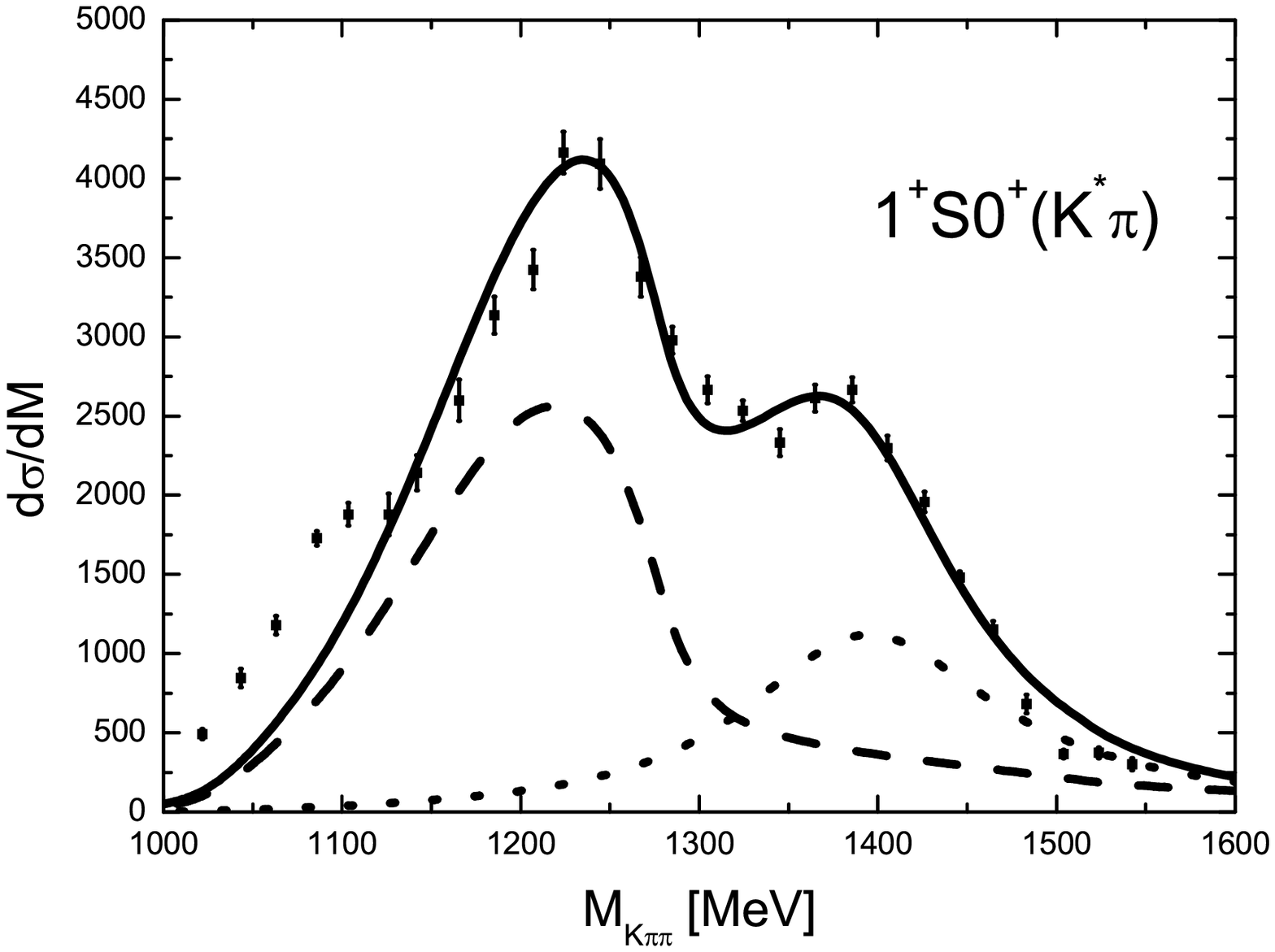} %
\includegraphics[scale=0.27]{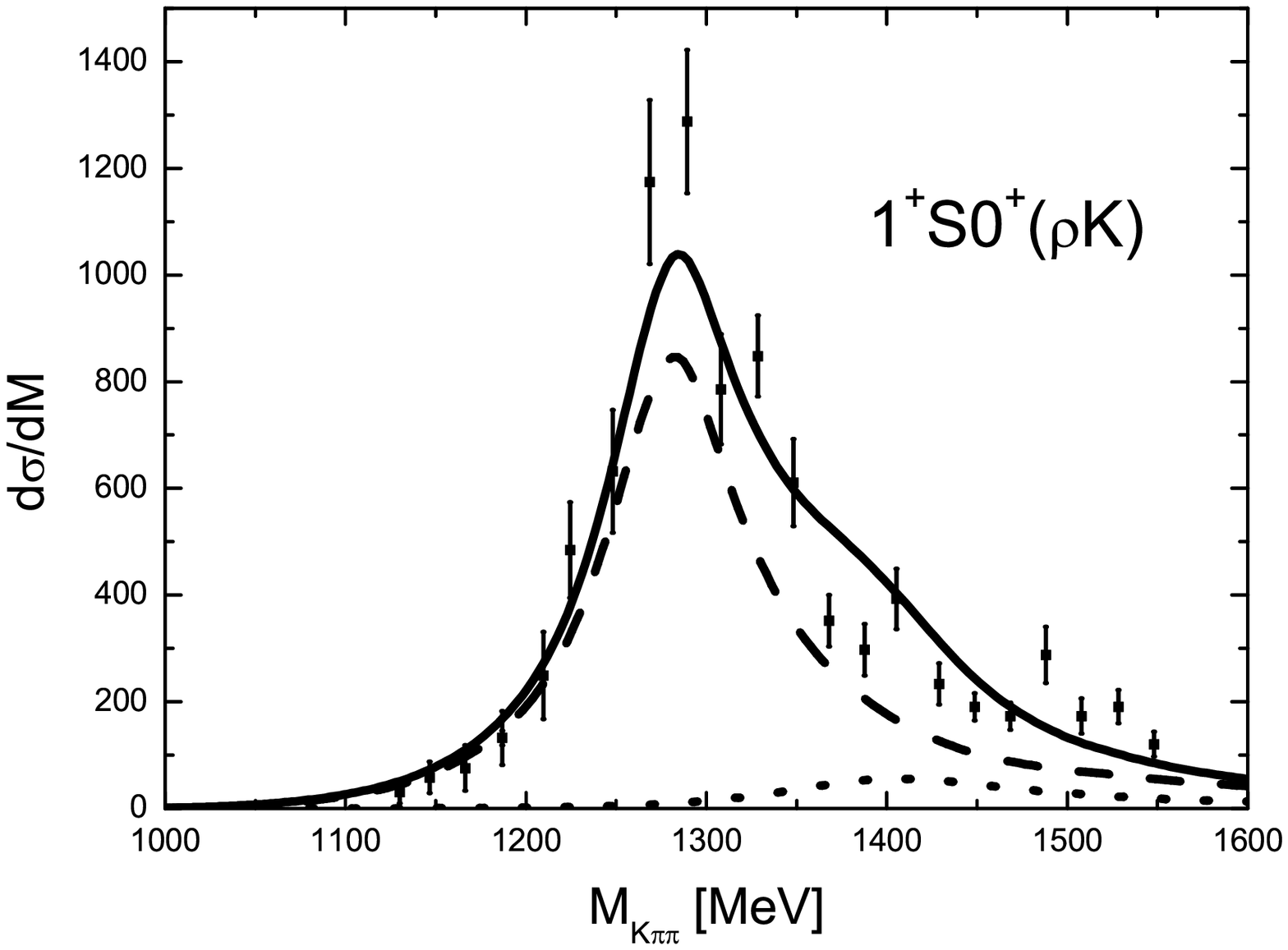}
\includegraphics[scale=0.27]{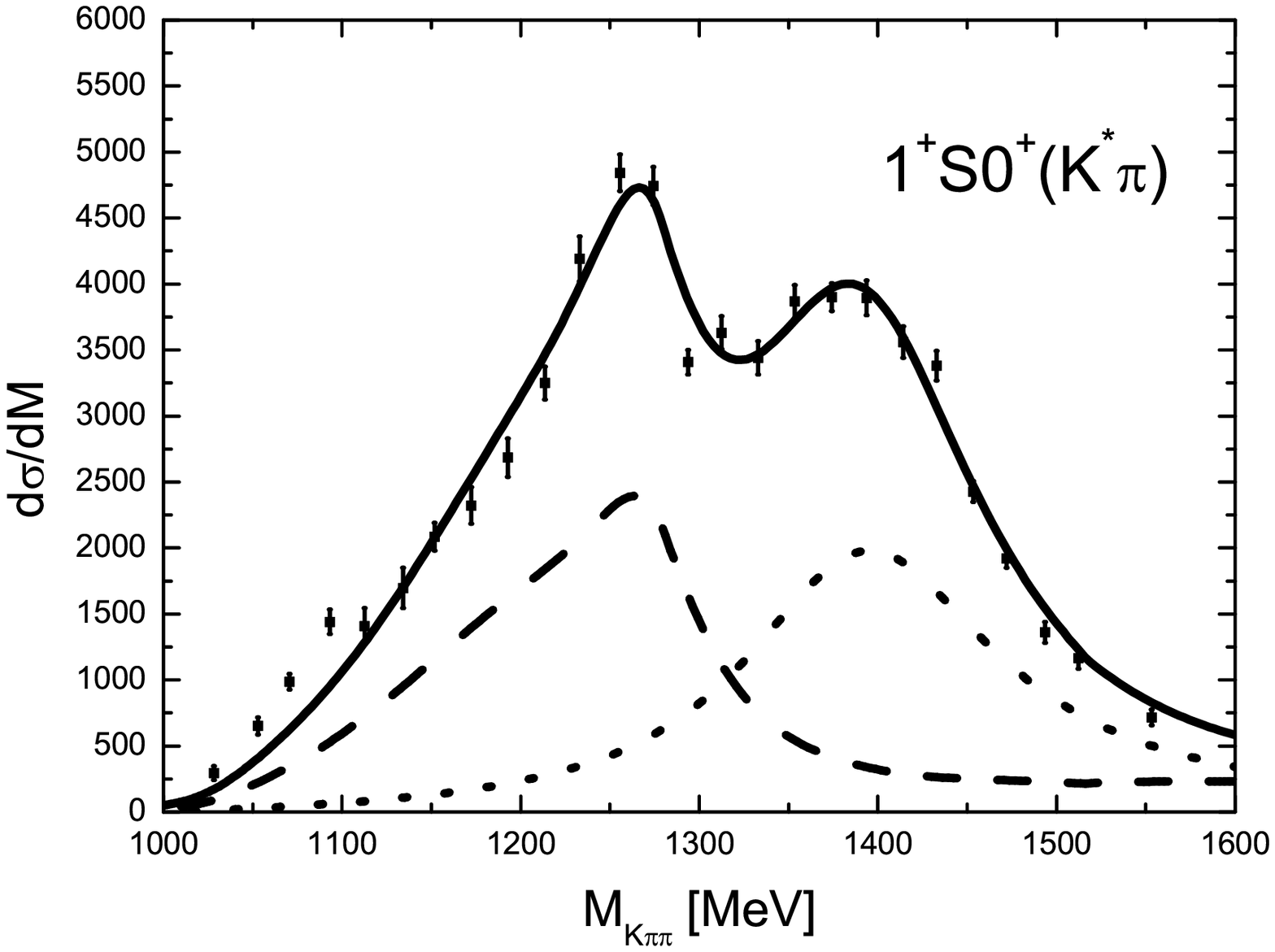}%
\includegraphics[scale=0.27]{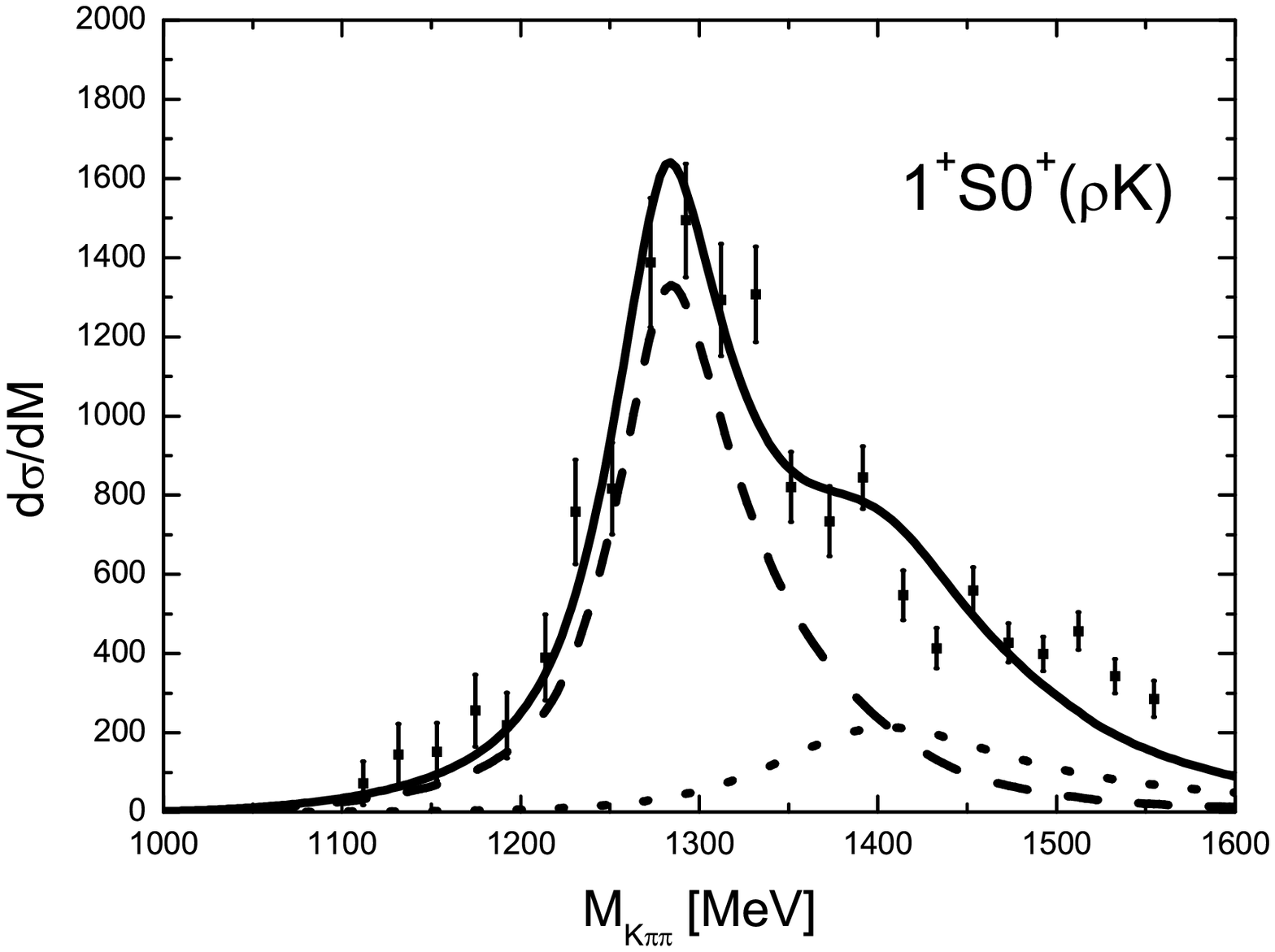}
\includegraphics[scale=0.27]{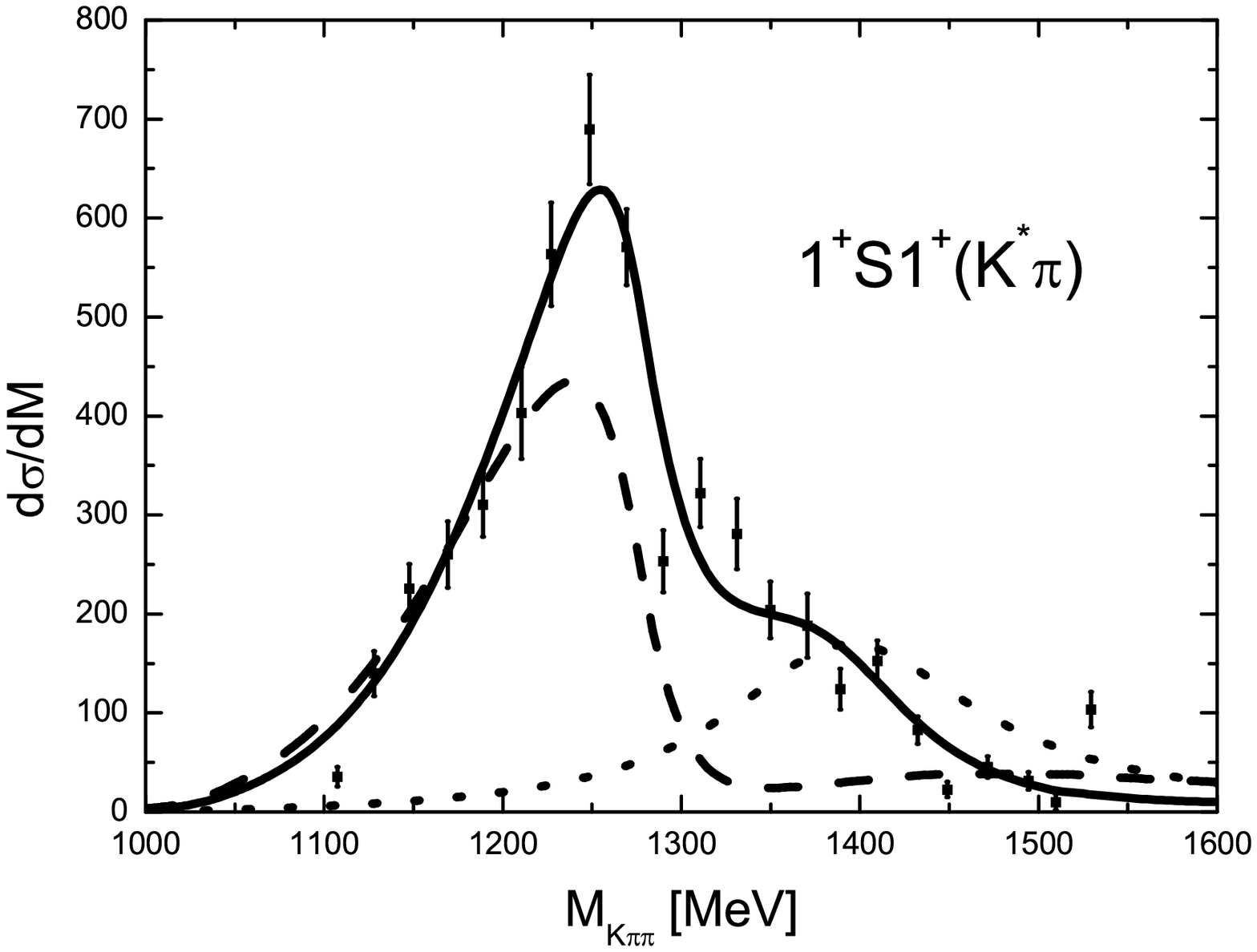}%
\includegraphics[scale=0.27]{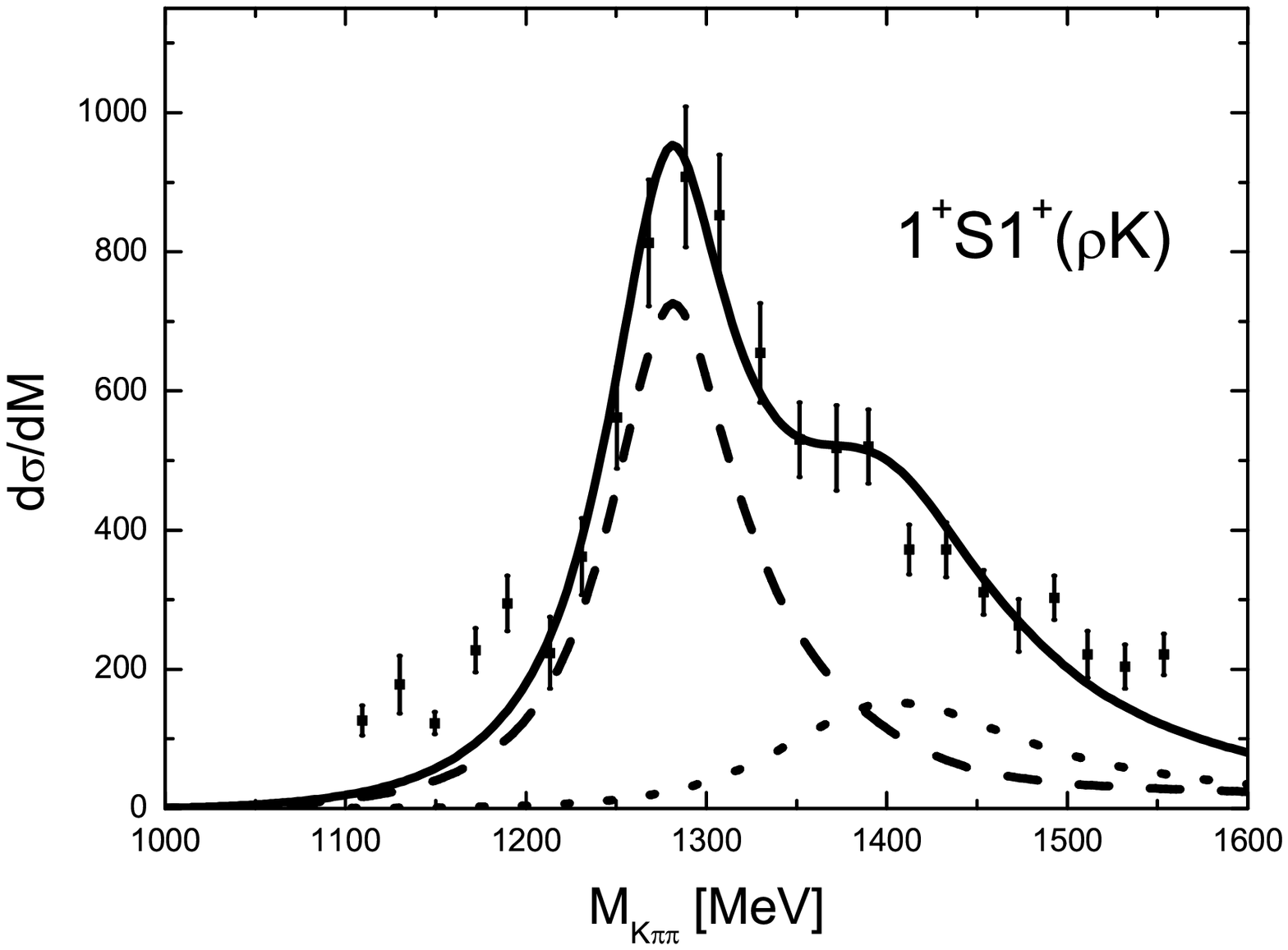}
\caption{$K^*\pi$ and $\rho K$ invariant mass distributions. The
data are from the WA3 reaction $K^- p\rightarrow K^-\pi^+\pi^- p$ at
63 GeV~\cite{Daum:1981hb}. Data in the upper panels are for $0\le
|t'|\le 0.05$ GeV$^2$ and those in the middle and bottom panels for
$0.05\le |t'|\le 0.7$ GeV$^2$, where $t'$ is the four momentum
transfer squared to the recoiling proton. The data are further
grouped by $J^PLM^\eta$ followed by the isobar and odd particle. $J$
is the total angular momentum, $P$ the parity, $L$ the orbital
angular momentum of the odd particle. $M^\eta$ denotes the magnetic
substate of the $K\pi\pi$ system and the naturality of the
exchange.}
\label{fig:s1}%fig3
\end{center}
\end{figure}
 According to Ref.~\cite{Flatte:1976xu},
for an $s$-wave resonance, the theoretical differential cross section
can be calculated by
 \begin{equation}\label{eq:inv}
 \frac{d\sigma}{dM}=c|T|^2q
 \end{equation}
 where $M$ is the invariant mass of the $K^*\pi$ or $\rho K$ systems,
 $c$ is a normalization constant, $T$ is the amplitude specified above for the $K^*\pi$ or $\rho K$
 channels
  and $q$ is the center of mass three-momentum  of $K^*\pi$ or $\rho K$. We have taken $c$ to be 1, or in other words, it has been
absorbed into the coupling constants $a$, $b$, $g_{K^*\pi}$ and
$g_{\rho K}$. The 
theoretical invariant mass distributions calculated with Eq.~( \ref{eq:inv}) 
are shown in Fig.~\ref{fig:s1} in comparison with the
WA3 data~\cite{Daum:1981hb}.

 From
Fig.~\ref{fig:s1},
 it is clearly seen that our model can fit the data around
the peaks very well. In Fig.~\ref{fig:s1},
 the dashed and dotted lines are the
separate contributions of the $K_1(1270)$ and the $K_1(1400)$. One can
easily see that the $K_1(1400)$ decays dominantly to $K^*\pi$, which is
consistent with our present understanding of this
resonance.

It should be mentioned that in our model the lower peak observed in
the invariant mass distribution of the $K^*\pi$ channel is due to
the contribution of the two poles of the $K_1(1270)$. This is very
different from the traditional interpretation. For example, the
lower peak observed in the $K^*\pi$ invariant mass distributions of
$K^\pm p\rightarrow K^\pm \pi^+\pi^- p$ at 13 GeV was interpreted
as a pure Gaussian background by Carnegie et
al.~\cite{Carnegie:1976cs}, which has a shape similar to the
contribution of the $K_1(1270)$ as shown in Fig.~\ref{fig:s1}.
 On the other
hand, the K-Matrix approach was adopted to analyze the WA3 data~\cite{Daum:1981hb}
and the SLAC data~\cite{Brandenburg:1975gv} .
In this latter approach, the lower peak mostly comes from the
so-called Deck background, which after unitarization, also has a
shape of resonance. As we mentioned in the introduction, even in the
original
 WA3 paper~\cite{Daum:1981hb}, it was noted that their model failed to
describe the $1^+S1^+$($K^*\pi$) data, in the notation $J^P L
M^\eta$ with $\eta$ the naturality of the
exchange~\cite{Daum:1981hb}. The predicted peak is 20 MeV higher
than the data. If the fit were done only to the $K^*\pi$ data, the
agreement was much better but then the predicted $K_1(1270)$ would
be lower by 35 MeV than that obtained when other channels were also
considered in the fit.

It is worth stressing that the $K_1(1270)$ peak seen in the
upper-left panel of Fig.~\ref{fig:s1}
 is significantly broader than that in the
upper-right panel. Furthermore the peak positions are also different
in the two cases (1240 MeV and 1280 MeV respectively). Both
features have a straightforward interpretation in our theoretical
description since the first one is dominated by the low-energy
(broader) $K_1(1270)$ state, while the second one is dominated by
the higher-energy (narrower) $K_1(1270)$ state.

\section{Summary and conclusion}

Studies based on unitary chiral perturbation theory obtain two poles
in the $I=1/2$, $S=1$, vector-pseudoscalar scattering amplitudes
which can be assigned to two $K_1(1270)$ resonances. One pole is at
$\sim1200$ MeV with a width of $\sim250$ MeV and the other is at
$\sim1280$ MeV with a width of $\sim150$ MeV. The lower pole
couples more to the $K^*\pi$ channel whereas the higher pole couples
dominantly to the $\rho K$ channel. Different reaction mechanisms
may prefer different channels and thus this  explains the
different invariant mass distributions seen in various experiments.

We have analyzed the WA3 data on the $K^-p\to K^-\pi^+\pi^-p$
reaction since it is the most conclusive and high-statistics
experiment quoted in the PDG on the $K_1(1270)$ resonance. Our model
obtains a good description of the WA3 data both for the $K^*\pi$ and
$\rho K$ final state channels. In our model,  the peak in the
$K\pi\pi$ mass distribution around the $1270$ MeV region is
a superposition of the two poles, but in the $K^*\pi$ channel the
lower pole dominates and in the $\rho K$ channel the higher pole
gives the biggest contribution.

\section{Acknowledgments}
This work is partly supported by
DGICYT Contract No. BFM2003-00856, FPA2004-03470, the Generalitat Valenciana, and the
E.U. FLAVIAnet network Contract No. HPRN-CT-2002-00311. This
research is part of the EU Integrated Infrastructure Initiative
Hadron Physics Project under Contract No. RII3-CT-2004-506078.

\end{document}